\documentclass{ws-ijmpcs}
\newcommand{\prd}{{\it Phys. Rev. D }}

\newcommand{\apj}{{\it Astrophys. J. }}

\newcommand{\ijmpd}{{\it Int. J. Mod. Phys. D }}

\newcommand{\mnras}{{\it Mon. Not. R. Astron. Soc. }}

\newcommand{\fNLl} {f_{\rm NL}^{\rm local}}

\def\simlt{\stackrel{<}{{}_\sim}}

\begin{document}

\markboth{Wilmar A. Cardona, Armando Bernui, Marcelo J. Rebou\c{c}as}
{A comparative study of non-Gaussianity in ILC--$7$yr}


%
\catchline{}{}{}{}{}
%

\title{A comparative study of non-Gaussianity in ILC--$7$yr CMB map}

\author{WILMAR A. CARDONA}
\address{Centro Brasileiro de Pesquisas F\'{\i}sicas, Rua Dr.\ Xavier Sigaud 150\\
22290-180 Rio de Janeiro -- RJ,  Brazil} 

\author{ARMANDO BERNUI}
\address{Instituto de Ci\^encias Exatas, Universidade Federal de Itajub\'a\\
37500-903 Itajub\'a -- MG, Brazil} 

\author{MARCELO J. REBOU\c{C}AS}  
\address{Centro Brasileiro de Pesquisas F\'{\i}sicas, Rua Dr.\ Xavier Sigaud 150\\
22290-180 Rio de Janeiro -- RJ,  Brazil} 

\maketitle

\begin{history}
\received{Day Month Year}
\revised{Day Month Year}
\end{history}

\begin{abstract}
A detection or non detection of primordial non--Gaussianity (NG) by using the cosmic microwave
background radiation (CMB) is a possible way to break the degeneracy of early universe models.
Since a single statistical estimator hardly can be sensitive to all possible forms of NG
which may be present in the data, it is important to use different statistical estimators
to study NG in CMB. Recently, two new large-angle NG indicators based on skewness and kurtosis
of spherical caps or spherical cells of CMB sky have been proposed and used in both CMB data
and simulated maps. Here, we make a comparative study of these two different procedures by
examining the NG in the WMAP seven years ILC map. We show that the spherical
cells procedure detects a higher level of NG than  that obtained by the method with
overlapping spherical caps.
\end{abstract}

\ccode{PACS numbers: 98.80.Es, 98.70.Vc, 98.80.-k}

\section{Introduction}	

Recent CMB cosmological observations are compatible with a nearly scale invariant
power spectrum. However, there are many models of early universe that fit CMB data.%
\cite{Inflation-reviews} This gives rise to the need of  further ways
of testing the models of primordial universe.  A possible way to break the degeneracy
in the models of the early universe is by studying deviation of Gaussianity
of CMB data.\cite{Komatsu-2010}\cdash\cite{Bartolo2004}
Thus, for example, in a single-field model of inflation, the amplitude $\fNLl$ of the
three-point correlation function counterpart in Fourier space --- the so-called bispectrum ---
can be calculated in terms of the slow-roll parameters and is very tiny
($ \fNLl \simlt 10^{-6} $).\cite{Gauss_Single-field}
Hence, any convincing detection of $\fNLl \gg 1 $  indicate a clear departure
from the slow-roll inflationary paradigm. On the other hand, if a no
significant $\fNLl$ is found from CMB observation, the standard
slow-roll scenario would clearly be favored.

In the study of  NG in the CMB data one ought to take into account that
there are contributions which do not have a primordial origin.
Some non-primordial contributions come from unsubtracted diffuse foreground emission,\cite{Bennett-etal-2003}\cdash\cite{Leach-etal-2008} unresolved
point sources,\cite{Komatsu-etal-2003} possible systematic errors,\cite{Su-Yadav-etal2010}
and secondary anisotropies such as gravitational weak lensing and the Sunyaev--Zeldovich effect.\cite{Komatsu-2010}\cdash\cite{PNG-rev-Liguori}
Deviation from Gaussianity may also have a cosmic topology origin
(see, e.g., the review articles Refs.~\refcite{CosmTopReviews}).
Different statistical tools can provide information from distinct contributions to the
NG in CMB data (see, for example, Ref.~\refcite{SomeNGrefs} and references therein).
Furthermore, since a single statistical estimator
hardly can be sensitive to all possible contributions to NG  that may exist, it is useful
to use different statistical tools to test CMB data for deviations from a Gaussian statistics
in order to quantify the amount of any non-Gaussian signals in the data, and extract information
on their possible origins.

Recently, two new large-angle NG indicators based on skewness and kurtosis
of spherical caps or spherical cells of CMB sky have been proposed and used
in both CMB data and simulated maps.\cite{Bernui-Reboucas2009}\cdash\cite{Bernui-Reboucas2012}
In this work,  we make a comparative study of these two different procedures by
examining the NG in the WMAP ILC--$ 7 $  map. We show that the spherical
cells procedure detects a higher level of NG than that obtained by the method with
overlapping spherical caps.

\section{Statistical estimators and Non--Gaussianity in ILC--$ 7 $ yr map}
\label{s1}

A simple way for describing deviation from a Gaussian distribution in CMB temperature
fluctuations is by calculating skewness
\begin{equation}
S = \frac{\mu_3}{\sigma^3}\, ,
\label{skewness definition}
\end{equation}
and kurtosis
\begin{equation}
K = \frac{\mu_4}{\sigma^4} - 3\, ,
\label{kurtosis definition}
\end{equation}
where $ \sigma $, $ \mu_3 $ and $ \mu_4 $ are, respectively, the second, third and fourth
central moments of the distribution. Based upon the fact that $S$ and $K$ vanish
for a Gaussian distribution, two statistical indicators to measure large-angle NG in CMB
were introduced in Ref.~\refcite{Bernui-Reboucas2009}.
The constructive process can be formalized as follows.
Let $\Omega_j \equiv \Omega(\theta_j,\phi_j) \in S^2$ be set of points in a spherical region.
For $ j=1,\, \dots,N $, we define scalar functions $ S : \Omega_j \rightarrow \mathbb{R}$
and $ K : \Omega_j \rightarrow \mathbb{R}$, that assign to the $j^{\,\rm{th}}$ spherical
region two real numbers given by
\begin{equation}
S_j   \equiv  \frac{1}{N_{\rm p} \,\sigma^3_{j} } \sum_{i=1}^{N_{\rm p}}
\left( T_i\, - \overline{T_j} \,\right)^3
\label{definition skewness}
\end{equation}
and
\begin{equation}
K_j   \equiv  \frac{1}{N_{\rm p} \,\sigma^4_{j} } \sum_{i=1}^{N_{\rm p}}
\left( T_i\, - \overline{T_j} \,\right)^4 - 3 \,,
\label{definition kurtosis}
\end{equation}
where $ N_{\rm p} $ is the number of pixels in the $ j^{\,\rm{th}} $ spherical region,
$ T_i $ is the temperature at the $i^{\,\rm{th}}$ pixel, $\overline{T_j}$ and
$\sigma_j$ are, respectively, the CMB mean temperature and the variance
\begin{equation}
\sigma_j^2 = \frac{1}{N_{\rm p}-1}\sum_{i=1}^{N_{\rm p}}(T_i - \overline{T_j})^2\, .
\end{equation}
in the $j^{\,\rm{th}}$ region.

In the next two section we shall use this constructive process to formalize two different
procedures which allow to build skewness and kurtosis $S(\theta,\phi)$ and $K(\theta,\phi)$
functions from a given input CMB map. The major difference of the two methods is
the way one chooses the spherical region to define these functions.

\subsection{Spherical caps method}
\label{ss1}

In this method one chooses for the spherical region, $\Omega_j$, overlapping spherical caps
in order to define $ S(\theta,\phi) $ and $ K(\theta,\phi) $ functions on the sphere.
The procedure to build these functions is as follows.

\begin{romanlist} 
\item
For a given CMB map we take a discrete set of points $ j=1,\, \dots,N_c $ homogeneously
distributed on the celestial sphere $S^2$ as the center of spherical caps
(spherical region $ \Omega_j $) with aperture $ \gamma $.
\item
Then, one calculates skewness ($ S_j $) and kurtosis ($ K_j $) for each spherical cap $ j $
defined, respectively, by Eqs.~\eqref{definition skewness}--\eqref{definition kurtosis}.
\item
Patching together the $S_j$ and $K_j$ values for each spherical cap we obtain two
discrete functions $ S(\theta,\phi) $ and $ K(\theta,\phi) $ defined on the celestial
sphere $ S^2 $. These functions can be used to measure NG as a function of the
angular coordinates $ (\theta,\phi) $. The Mollweid projection of skewness and kurtosis
functions $ S=S(\theta,\phi) $ and $ K=K(\theta,\phi) $ are skewness and kurtosis maps
which we shall refer hereafter as $S$ map  and $K$ map, respectively.
\end{romanlist}

\begin{figure}[h!]
\centering
\includegraphics[scale=0.24, angle=90]{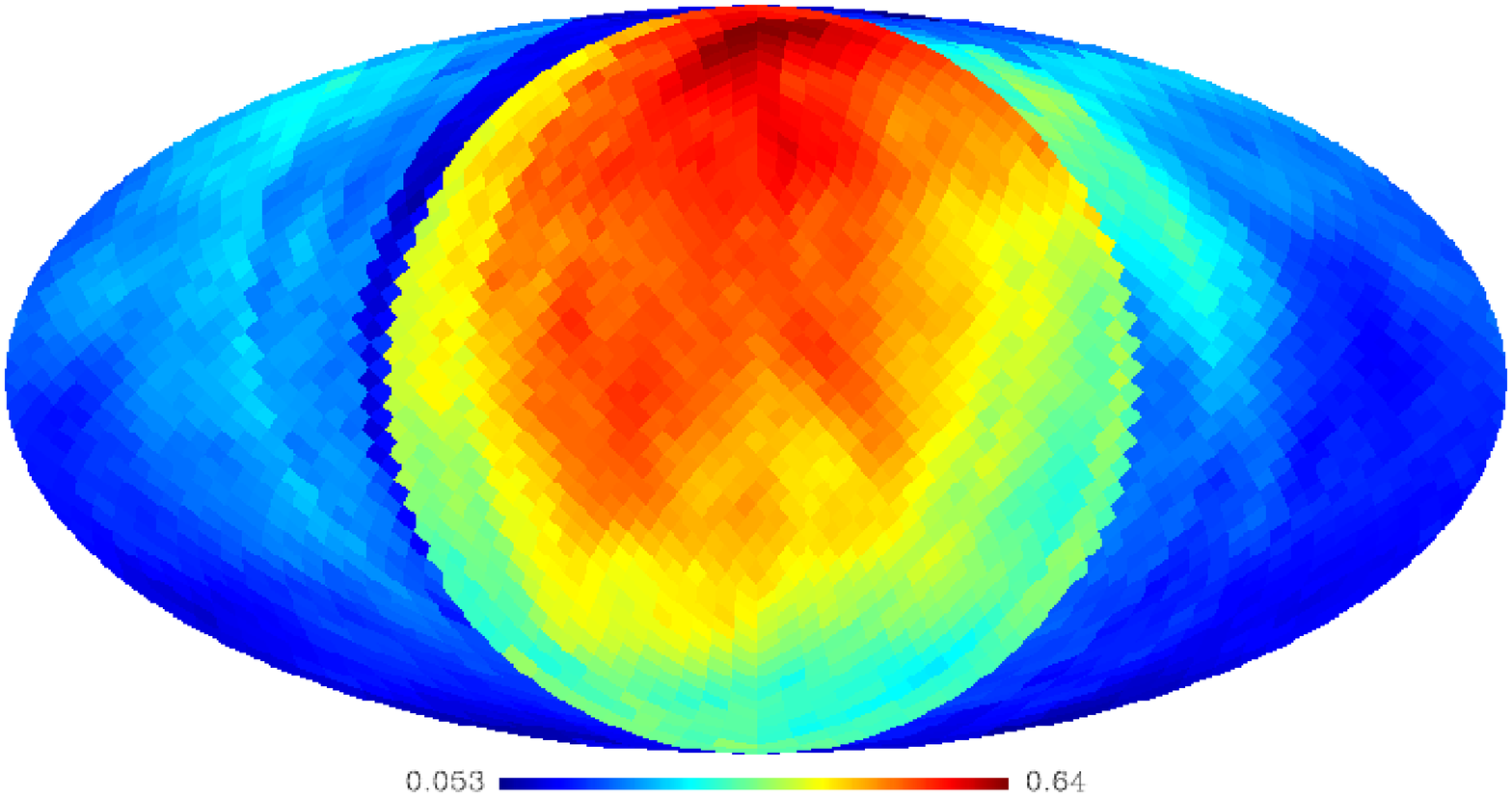}
\includegraphics[scale=0.24, angle=90]{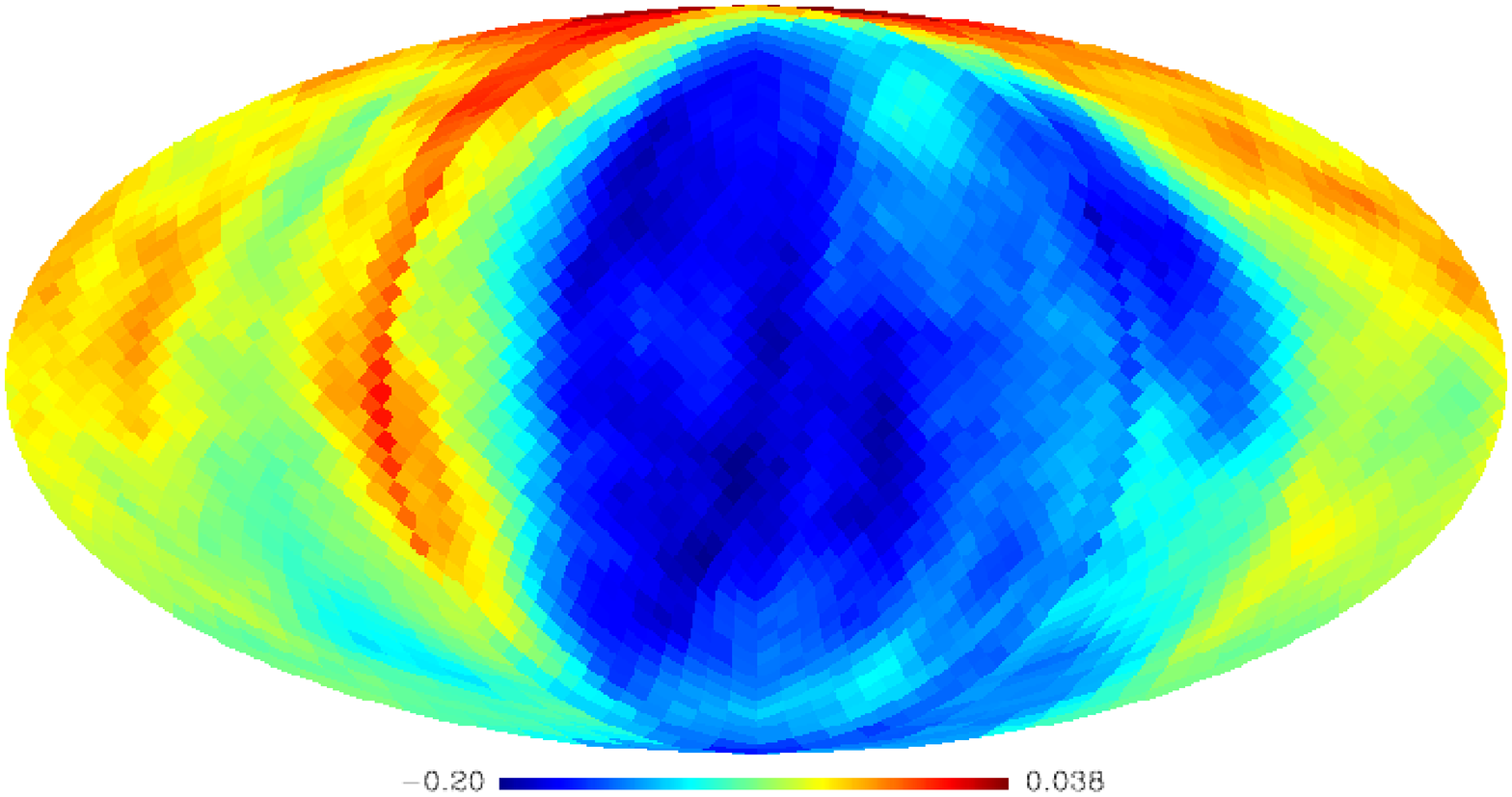}
\caption{Kurtosis indicator map (left panel) and skewness indicator map (right panel)
generated from the WMAP seven years ILC map by using the spherical caps method with
$ N_c = 3\,072 $.}
\label{caps method maps}
\end{figure}

Fig.~\ref{caps method maps} shows $S$ and $K$ maps calculated from the WMAP ILC-$7$ yr
map by using the spherical cap procedure.%
\footnote{In this work all the CMB maps we use to generate $S$ and $K$  maps have
HEALPix parameter $ N_{\rm side}=256 $ as defined in Ref.~\refcite{Gorski-et-al-2005}.
This means that each spherical cap has $ N_{\rm p} = 393\,216 $ pixels.
On the other hand, all $S$ and $K$ maps generated with the spherical caps
method have $N_c = 3\,072$ and $\gamma = 90^{\circ}$.}

\subsection{Spherical cells method}
\label{ss2}

Differently from the spherical caps method of the previous section, in the spherical cells method
one considers non-overlapping region of the CMB sphere, generated by the HEALPix partition of
the sphere, for defining $S(\theta,\phi)$ and $K(\theta,\phi)$ functions. The procedure to
build the indicators is as follows.

\begin{romanlist}
\item
For a given CMB map we divide the CMB sphere into $12$ equal area primary
spherical cells by using the HEALPix partition of the sphere.
\item
We divide each one of the $12$ primary cells, in which the HEALPix partition divides
initially the sphere, in $ N^{\prime\,2}_{\rm side} $ new spherical cells.
Thus, we obtain $ N^{\prime}_{\rm p} = 12 \times  N^{\prime\,2}_{\rm side}$ spherical
cells on the whole sphere.
\item
To define $ S(\theta,\phi) $ and $ K(\theta,\phi) $ functions on the sphere,
in each one of the $ N^{\prime}_{\rm p}  $ spherical cells we calculate skewness
and kurtosis by using Eqs.~\eqref{definition skewness}--\eqref{definition kurtosis}
with $ N_{\rm p} $ and $ \sigma_j $ being, respectively, the number of pixels and
the variance in each cell.
\item
Finally, patching together $ S_j $ and $ K_j $ ($ j=1,\, 2,\, \dots,\,  N^{\prime}_{\rm p}$)
we define discrete functions of skewness $ S(\theta,\phi) $ and kurtosis $ K(\theta,\phi) $
on the celestial sphere $ S^2 $. As in the spherical caps method, the Mollweid projection of
these functions constitutes $S$ and $K$ maps, respectively.
\end{romanlist}

\begin{figure}[h!]
\centering
\includegraphics[scale=0.24, angle=90]{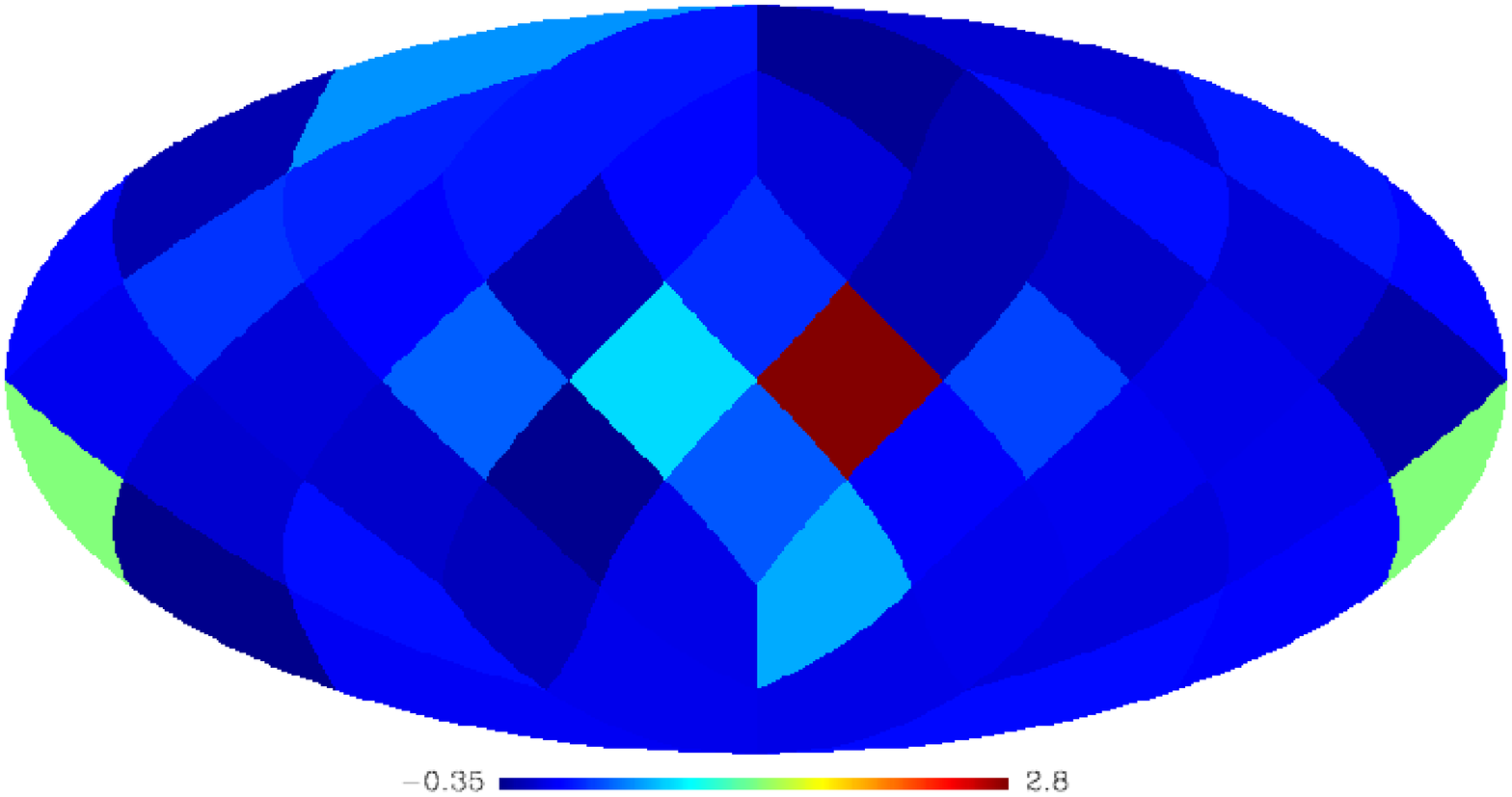}
\includegraphics[scale=0.24,angle=90]{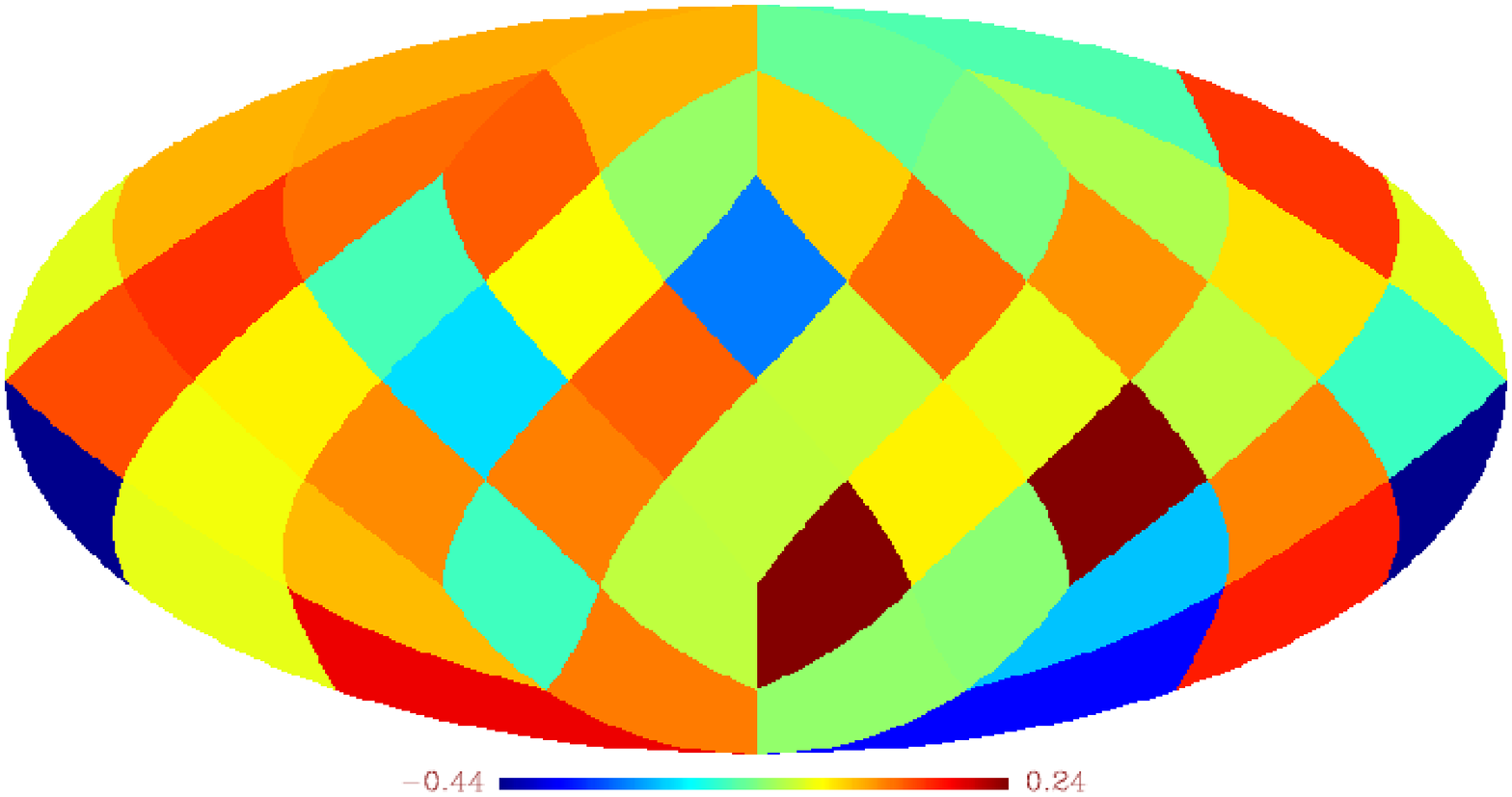}
\caption{Kurtosis indicator map (left panel) and skewness indicator map (right panel)
generated from the WMAP seven years ILC map by using the spherical cells method with
$ 48 $ spherical cells ($N^{\prime\,2}_{\rm side}=4$). Each colored division is a
spherical cell.}
\label{K map 2}
\end{figure}

Fig.~\ref{K map 2} shows $S$ and $K$ maps calculated from the WMAP ILC--$ 7 $ yr
by taking $48$ spherical cells each one with $ N_{\rm p} = 16\,384 $ temperature
fluctuation pixels.

\subsection{Comparative analysis}

In this section we shall make a comparative study of these two different procedure to
construct the $S$ and $K$ indicators by examining the NG in the WMAP  ILC-$7$ yr
map.
To this end, we first compute angular power spectra $ S_\ell $ and $ K_\ell $ of $S$ and $K$
maps generated from both a set of $1\,000$ Gaussian CMB simulated maps and from the WMAP
seven years ILC map.
Then,  we use  $ \chi^2$ statistics  to compare the sets of mean angular power spectra
$ \overline{S^{\rm G}_\ell} $ and $ \overline{K^{\rm G}_\ell} $ of  $S$ and $K$
Gaussian simulated CMB maps, with the angular power spectra, $ S^{\rm ILC}_{\ell} $ and
$ K^{\rm ILC}_{\ell} $ calculated from $S$ and $K$ maps generated from the WMAP ILC-$7$ yr
map. The procedure is as follows.


\begin{romanlist}[(ii)]
\item
We use $1\,000$ Gaussian CMB simulated maps which were built by using the procedure
explained in Ref. \refcite{ElsnerWandelt2009}.%
\footnote{These maps were made available to download in
http://planck.mpa--garching.mpg.de/cmb/fnl--simulations.}
\item
We generate $S$ and $K$ maps both for the set of $1\,000$ Gaussian CMB simulated maps
and for the ILC--$ 7 $ yr map.
\item
We expand the functions $ S(\theta,\phi) $ and $ K(\theta,\phi) $ corresponding to
the $S$ and $K$ maps of the item $ (ii) $ in spherical harmonics as

\begin{eqnarray}
K (\theta,\phi) & = & \sum_{\ell=0}^\infty \sum_{m=-\ell}^{\ell} b_{\ell m}
\,Y_{\ell m} (\theta,\phi) \, , \nonumber \\
S (\theta,\phi) & = & \sum_{\ell=0}^\infty \sum_{m=-\ell}^{\ell} b^{'}_{\ell m}
\,Y_{\ell m} (\theta,\phi)\, ,
\label{expansion in harmonics for kurtosis function}
\end{eqnarray}

and find the corresponding angular power spectra given by

\begin{eqnarray}
K_{\ell} & = & \frac{1}{2\ell+1} \sum_m |b_{\ell m}|^2 \, , \nonumber \\
S_{\ell} & = & \frac{1}{2\ell+1} \sum_m |b^{'}_{\ell m}|^2 \, .
\label{angular power spectrum}
\end{eqnarray}

\item
For the sets of Gaussian CMB simulated maps we also calculate the mean value
$ \overline{S_\ell} $ and $ \overline{K_\ell} $ and their variance.
\item
Finally, we calculate $ \chi^2 $ for $S$ and $K$ maps given by

\begin{equation}
\chi^2_{S_\ell} = \frac{1}{3}\sum_{\ell=1}^4 \frac{\left( S^{\rm ILC}_\ell
-\overline{S^{\rm G}_\ell}\right)^2}{{\sigma^{\rm G}_\ell}^2}\, ,
\label{chi quadrado}
\end{equation}
and
\begin{equation}
\chi^2_{K_\ell} = \frac{1}{3}\sum_{\ell=1}^4 \frac{\left( K^{\rm ILC}_\ell
-\overline{K^{\rm G}_\ell}\right)^2}{{\sigma^{\rm G}_\ell}^2}\, .
\label{chi quadrado}
\end{equation}
\end{romanlist}
\begin{figure}[h!]
\centering
\includegraphics[scale=0.35]{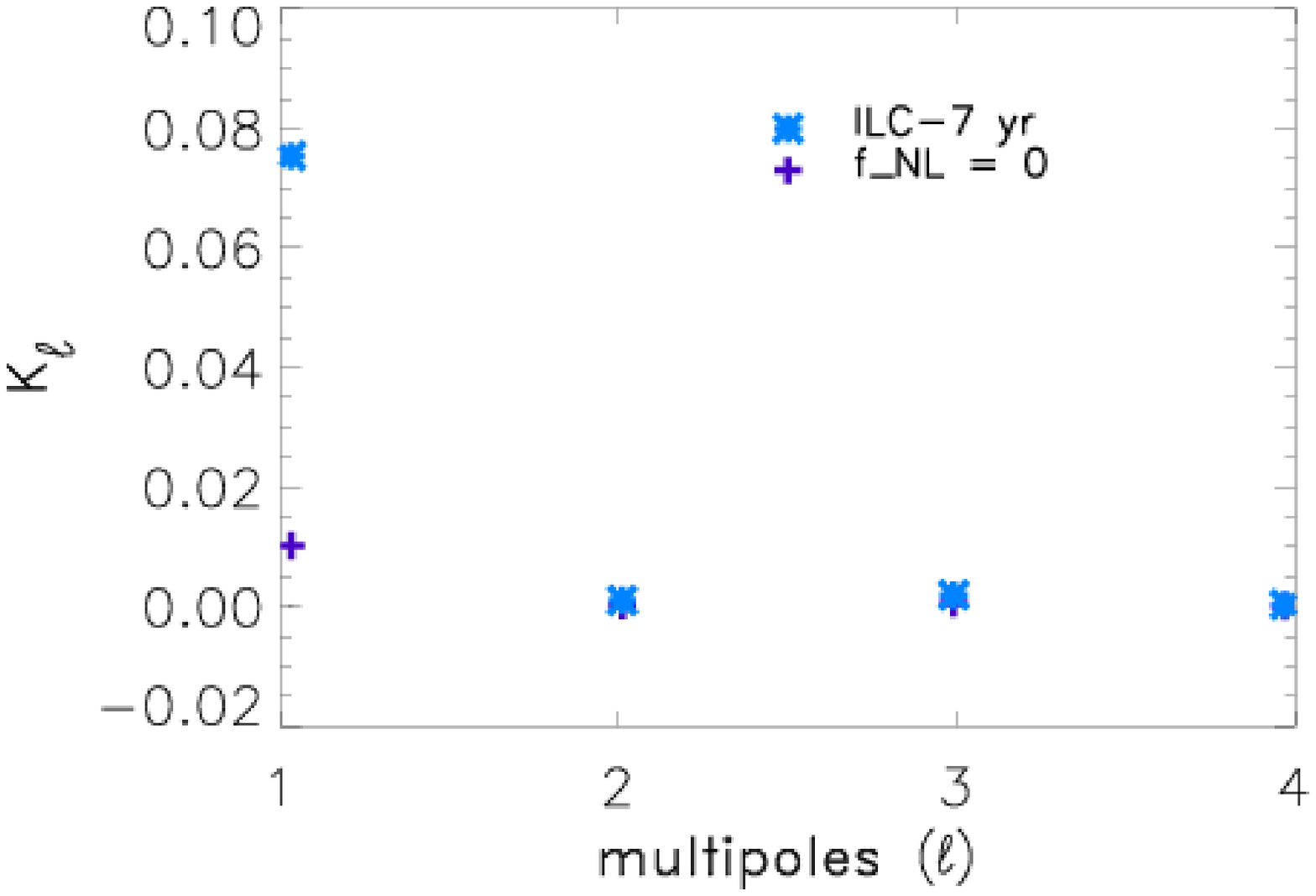}
\includegraphics[scale=0.35]{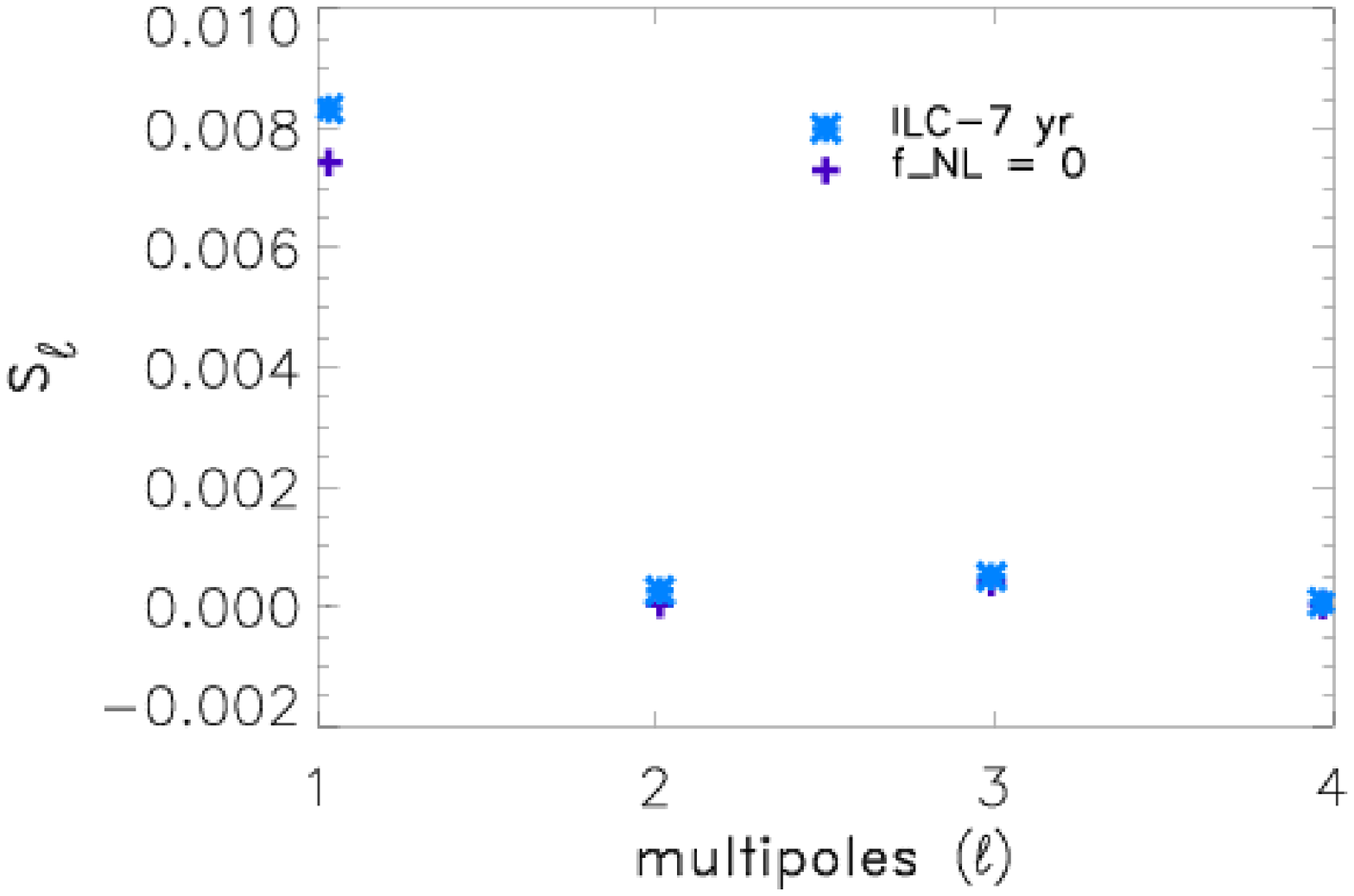}
\caption{Low $ \ell $ angular power spectrum of skewness $S_\ell $ (right panel) and kurtosis
$ K_\ell $ (left panel) maps calculated from ILC--$ 7 $ yr and Gaussian CMB simulated maps
by using the spherical caps method.}
\label{fig1}
\end{figure}

\begin{table}[h!]
\tbl{$ \chi^2 $ test goodness of fit for $ S^{\rm ILC}_\ell $ and $ K^{\rm ILC}_\ell $
as compared with $ \overline{S^{\rm G}_\ell} $  and $ \overline{K^{\rm G}_\ell} $.}
{\begin{tabular}{|c|c|c|}
\hline
Method & $ \chi^2_{S_\ell} $ & $\chi^2_{K_\ell}$ \\ \hline
Cells     &  $ 6.27\times 10 $  & $ 6.25\times10^3 $ \\ \hline
Caps     &  $ 4.07 $                &  $ 2.04\times10 $    \\ \hline
\end{tabular}
\label{table 1}}
\end{table}

\begin{figure}[h!]
\centering
\includegraphics[scale=0.35]{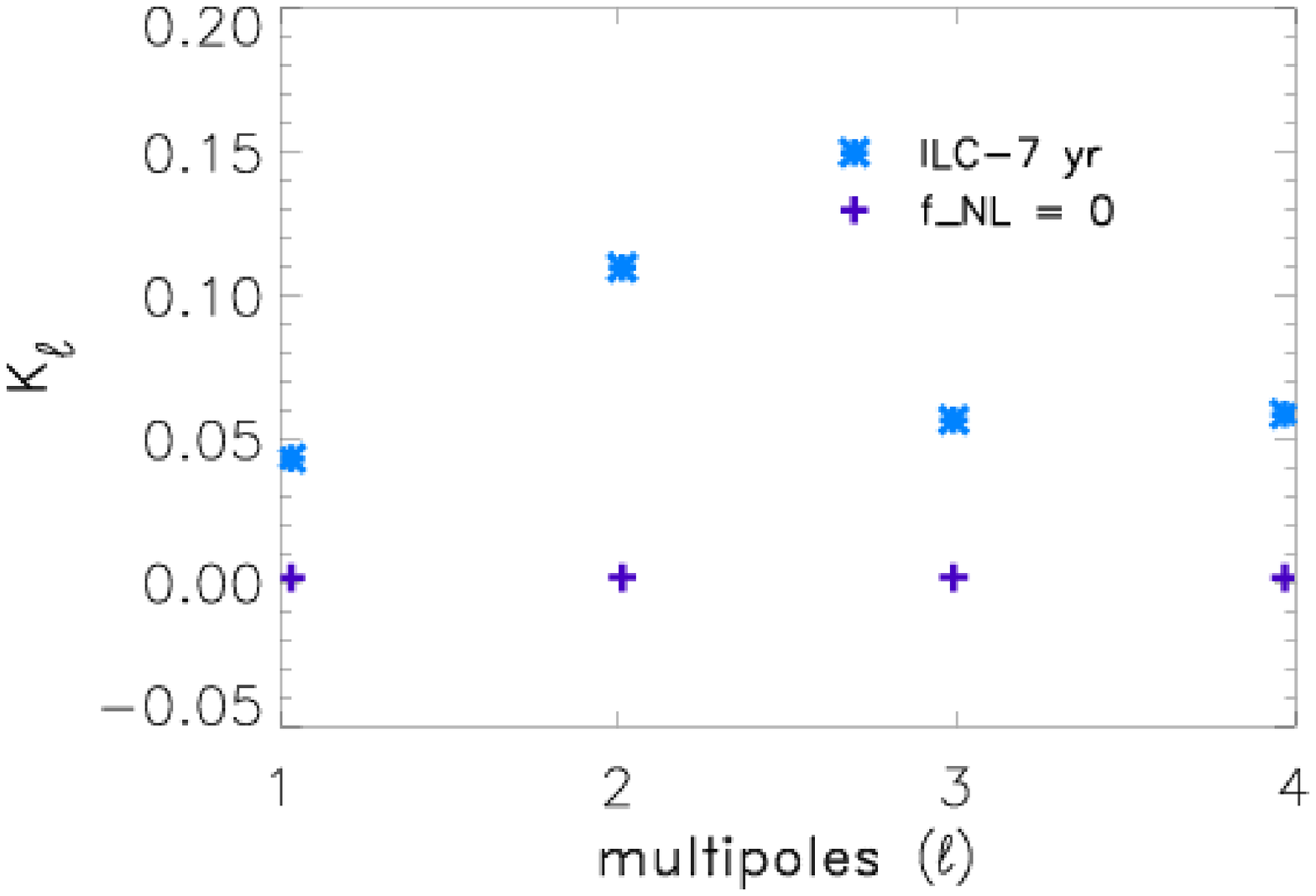}
\includegraphics[scale=0.35]{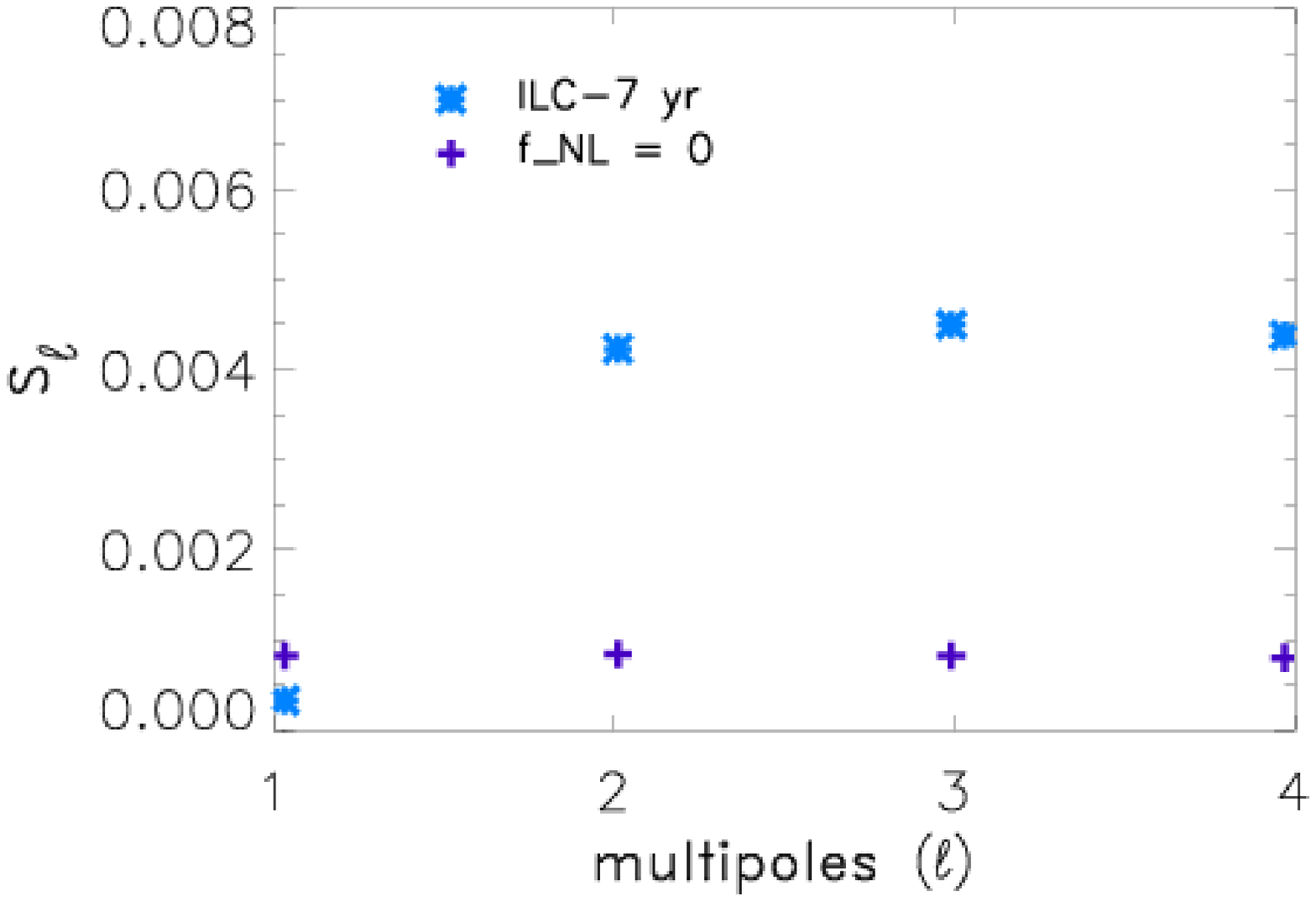}
\caption{Low $ \ell $ angular power spectrum of skewness $ S_\ell $ (right panel) and kurtosis
$ K_\ell $ (left panel) maps calculated from ILC--$ 7 $ yr and Gaussian CMB simulated maps
by using the spherical cells method.}
\label{fig2}
\end{figure}

Fig.~\ref{fig1} and Fig.~\ref{fig2} show the angular power spectra for $S$ and $K$ maps
calculated by using the spherical caps method and  the cells method, respectively.
Fig.\ref{fig1} makes apparent the tiny difference in the power of the spectra
$ \overline{S^{\rm G}_\ell} $  ($ \overline{K^{\rm G}_\ell} $) and $ S^{\rm ILC}_\ell $
($ K^{\rm ILC}_\ell $).
On the other hand, Fig.~\ref{fig2} exhibits bigger differences between
$ \overline{S^{\rm G}_\ell} $  ($ \overline{K^{\rm G}_\ell} $) and $ S^{\rm ILC}_\ell $
($ K^{\rm ILC}_\ell $)
making showing qualitatively that the spherical cells method capture a greater departure
for the ILC--$ 7 $ yr map than that detected through the spherical method.



To obtain quantitative information regarding the sensitivity to detect deviation
from Gaussianity by $S$ and $K$ indicators built by using both spherical caps method
and spherical cells method, we perform a $ \chi^2 $ test to determine the goodness
of fit for $ S^{\rm ILC}_\ell $ ($ K^{\rm ILC}_\ell $) as compared with
$ \overline{S^{\rm G}_\ell} $  ($ \overline{K^{\rm G}_\ell} $).
Since a good fit occurs when $ \chi^2 \approx 1 $, results collected together
in Table~\ref{table 1} make apparent that, for the ILC--$ 7 $ yr map, the
spherical cells method capture a greater degree of NG than that detected
through the spherical method.


\section*{Acknowledgments}
M.J. Rebou\c{c}as acknowledges the support of FAPERJ under a CNE E-26/101.556/2010 grant.
This work was also supported by Conselho Nacional de Desenvolvimento
Cient\'{\i}fico e Tecnol\'{o}gico (CNPq) - Brasil, under grant No. 475262/2010-7.
A. Bernui thanks FAPEMIG for the grant APQ--01893--10.
M.J. Rebou\c{c}as, Wilmar A. Cardona and A. Bernui thank  CNPq for the grants under
which this work was carried out. Some of the results in this paper were derived using the
HEALPix package.\cite{Gorski-et-al-2005}



\begin{thebibliography}{00}  

\bibitem{Inflation-reviews} B.A. Bassett, S. Tsujikawa, and D. Wands,
{\it Rev. Mod. Phys.} \textbf{78}, 537 (2006);
A. Linde, {\it Lect. Notes Phys.} \textbf{738}, 1 (2008).

\bibitem{Komatsu-2010} E. Komatsu, {\it Class. Quant. Grav.} \textbf{27}, 124010 (2010)

\bibitem{Bartolo2004}
N. Bartolo, E. Komatsu, S. Matarrese, and A. Riotto,
{\it Phys. Rept.} {\bf 402}, 103 (2004). 

\bibitem{Gauss_Single-field}
V. Acquaviva, N. Bartolo, S. Matarrese, and A. Riotto,
Nucl. Phys. B \textbf{667}, 119 (2003);
J. Maldacena, JHEP \textbf{0305} 013 (2003);  
M. Liguori, F.K. Hansen, E. Komatsu, S. Matarrese, and A. Riotto,
\prd \textbf{73}, 043505 (2006).

\bibitem{Bennett-etal-2003} C.L. Bennett \emph{et al.}, Astrophys. J. Suppl. \textbf{148},
97 (2003).

\bibitem{Leach-etal-2008} S.M. Leach \emph{et al.}, Astron. Astrophys. \textbf{491}, 597 (2008).

\bibitem{Komatsu-etal-2003} E. Komatsu \emph{et al.}, Astrophys. J. Suppl. \textbf{148}, 119 (2003).

\bibitem{Su-Yadav-etal2010}  M. Su, A.P.S. Yadav, M. Shimon, B.G. Keating,
\prd \textbf{83}, 103007 (2011). 

\bibitem{PNG-rev-Liguori} M. Liguori, E. Sefusatti, J.R. Fergusson, and E.P.S. Shellard,
{\it Adv. Astron.} \textbf{2010}, 980523 (2010).

\bibitem{CosmTopReviews}
G.F.R. Ellis, \textit{Gen. Rel. Grav.} \textbf{2}, 7 (1971);
M. Lachi\`{e}ze-Rey and J.-P. Luminet, \textit{Phys. Rep.} \textbf{254}, 135 (1995);
G.D. Starkman, \textit{Class. Quantum Grav.} \textbf{15},  2529 (1998);
J.-P. Uzan, R. Lehoucq, and J-P. Luminet, arXiv:gr-qc/0005128;
J. Levin, \textit{Phys. Rep.} \textbf{365},  251 (2002);
M.J. Rebou\c{c}as and G. I. Gomero, \textit{Braz. J. Phys.} \textbf{34}, 1358 (2004);
M.J. Rebou\c{c}as, arXiv:astro-ph/0504365;
M.C. Bento, O. Bertolami, M.J. Reboucas, P.T. Silva,  \textit{Phys. Rev. D } \textbf{73}, 043504 (2006). 

\bibitem{Bernui-Reboucas2009} A. Bernui and M.J. Rebou\c{c}as, \prd
\textbf{79}, 063528 (2009).

\bibitem{Bernui-Reboucas2010} A. Bernui and M.J. Rebou\c{c}as, \prd
\textbf{81}, 063533 (2010). 
%
\bibitem{Bernui-Reboucas2012} A. Bernui and M.J. Rebou\c{c}as, \prd
\textbf{85}, 023522 (2012). 

\bibitem{Cardona-Bernui-Reboucas2012} W.A. Cardona, A. Bernui, and
M.J. Rebou\c{c}as,  in preparation (2013).

\bibitem{SomeNGrefs}
A. Bernui, B. Mota, M.J. Rebou\c{c}as, and R. Tavakol,
{\it Astron. \&  Astrophys.} \textbf{464}, 479 (2007);
A. Bernui, B. Mota, M.J. Rebou\c{c}as, and R. Tavakol,
\ijmpd \textbf{16}, 411 (2007).
R. Saha, {\it Astrophys. J. Letters}, \textbf{739}, L56 (2011);
C. R\"ath, G.E. Morfill, G. Rossmanith, A.J. Banday, and K.M. G\'orski,
{\it Phys. Rev. Lett.} \textbf{102}, 131301 (2009);
G. Rossmanith, C. R\"ath, A.J. Banday, and G. Morfill,
\mnras \textbf{399}, 1921 (2009);
C. R\"ath, P. Schuecker, and A.J. Banday,
\mnras \textbf{380}, 466 (2007);
%
N. Mandolesi, C. Burigana, A. Gruppuso, and P. Natoli.
{\it J. Phys. Conf. Ser.} \textbf{335}, 012009 (2011);
%
A. Gruppuso, F. Finelli, P. Natoli, F. Paci, P. Cabella, A. De Rosa, and N. Mandolesi,
\mnras \textbf{411}, 1445 (2011),
%
S.-Y. Zhong, X. Wu, S.-Q. Liu, and X.-F. Deng,
\prd \textbf{82}, 124040 (2010);
S.-Y. Zhong and X. Wu,
\prd \textbf{81}, 104037 (2010);
A. Bernui, M.J. Rebou\c{c}as, and A.F.F. Teixeira,
arXiv:1005.0883 [astro-ph.CO];
A. Bernui, M.J. Rebou\c{c}as, and A.F.F. Teixeira,
{\it Int. J. Mod. Phys. Conf. Ser.} \textbf{3}, 286 (2011);
V.N. Yershov, V.V. Orlov, and A.A. Raikov,
arXiv:1205.5139 [astro-ph.CO].

\bibitem{Gorski-et-al-2005} K.M. G\'orski, E. Hivon, A.J. Banday, B.D. Wandelt,
F.K. Hansen, M. Reinecke, and M. Bartelman, \apj \textbf{622}, 759 (2005).


\end{thebibliography}
\end{document}